\documentclass[a4paper,prl,twocolumn,showpacs,superscriptaddress,floatfix,nofootinbib]{revtex4-1}
\usepackage{lipsum} \usepackage{graphicx} \usepackage{epsf} \usepackage{epsfig}
\usepackage{amssymb,amsmath,amsfonts} \usepackage[usenames]{color}
\usepackage{amssymb} \usepackage{times} \usepackage{bm}
\usepackage{pdfpages}
\makeatletter
\AtBeginDocument{\let\LS@rot\@undefined}
\makeatother
\usepackage{hyperref} \hypersetup{
  colorlinks=true,        
  linkcolor=blue,         
  citecolor=cyan,         
  }

\usepackage{graphicx}  
\usepackage{dcolumn}   
\usepackage{bm}        
\usepackage{amsfonts,amsmath,amssymb,mathrsfs} 
\usepackage{color}

\usepackage{hyperref} \hypersetup{
  colorlinks=true,        
  linkcolor=blue,         
  citecolor=cyan,         
  } 
\usepackage{natbib,times}

\definecolor{orange}{rgb}{1,0.5,0} \definecolor{dgreen}{rgb}{0.0, 0.5, 0.0}

\newcommand{\cf}{cf.~} \newcommand{\ie}{i.e.,~} \newcommand{\eg}{e.g.,~}

\renewcommand{\BibitemShut}[1]{}

\begin{document}

\title{Postmerger Gravitational-Wave Signatures of Phase Transitions in Binary
Mergers}

\author{Lukas~R.~Weih}
\affiliation{Institut f{\"u}r Theoretische Physik,
  Max-von-Laue-Stra{\ss}e 1, 60438 Frankfurt, Germany}
\author{Matthias~Hanauske} 
\affiliation{Institut f{\"u}r Theoretische Physik,
  Max-von-Laue-Stra{\ss}e 1, 60438 Frankfurt, Germany}
\affiliation{Frankfurt Institute for Advanced Studies,
  Ruth-Moufang-Stra{\ss}e 1, 60438 Frankfurt, Germany}

\author{Luciano~Rezzolla}
\affiliation{Institut f{\"u}r Theoretische Physik,
  Max-von-Laue-Stra{\ss}e 1, 60438 Frankfurt, Germany}
\affiliation{School of Mathematics, Trinity College, Dublin 2, Ireland}

\begin{abstract} 
With the first detection of gravitational waves from a binary system of
neutron stars, GW170817, a new window was opened to study the
properties of matter at and above nuclear-saturation density. Reaching
densities a few times that of nuclear matter and temperatures up to
$100\,\rm{MeV}$, such mergers also represent potential sites for a phase
transition (PT) from confined hadronic matter to deconfined quark matter.
While the lack of a postmerger signal in GW170817 has prevented us from
assessing experimentally this scenario, two theoretical studies have
explored the postmerger gravitational-wave signatures of PTs in
mergers of binary systems of neutron stars.
We here extend and complete the picture by presenting a novel
signature of the occurrence of a PT. More specifically, using fully
general-relativistic hydrodynamic simulations and employing a suitably
constructed equation of state that includes a PT, we present the
occurrence of a ``delayed PT'', \ie a PT that develops only some time
after the merger and produces a metastable object with a quark-matter
core, \ie a hypermassive hybrid star. Because in this scenario, the
postmerger signal exhibits two distinct fundamental gravitational-wave
frequencies -- before and after the PT -- the associated signature
promises to be the strongest and cleanest among those considered so far,
and one of the best signatures of the production of quark matter in the
present Universe.
\end{abstract}


\maketitle


\noindent\emph{Introduction.~}The first detection of gravitational waves
from a merging binary system of neutron stars (BNS) \cite{Abbott2017},
GW170817, and of its electromagnetic counterpart
\cite{Abbott2017b} has provided a wealth of information
not only on the nature of gravity, but also on the properties of the
equation of state (EOS) of nuclear matter
\cite{Margalit2017,Bauswein2017b,Rezzolla2017,Ruiz2017,Annala2017,Radice2017b,Most2018,Shibata2019,Koeppel2019}.
The understanding of all the information extracted from this event was
aided by numerical simulations that predicted the properties of the
gravitational-wave signal
\cite{Bauswein2011,Takami:2014,Takami2015,Bernuzzi2015a,Rezzolla2016,
  Maione2017,Kawaguchi2018,Dietrich2019} and the kilonova resulting from
radioactive decay of heavy elements that are produced via $r$ process in
the merger's ejected material
\cite{Dietrich2016,Siegel2016a,Bovard2017,Perego2017,Fujibayashi2017b,Siegel2017,Fernandez2018}.
These simulations have shown that after the inspiral and merger, a
hypermassive neutron star (HMNS) is formed. The fate of the HMNS depends
on a number of factors, such as the mass, mass ratio, strength of
the magnetic field, and, of course, the underlying EOS. An important
degree of freedom associated with the EOS is the possibility of a phase
transition (PT) from hadronic to quark matter. Indeed, considering the
high densities in neutron-star cores (up to $\sim(6-7)\,\rho_0$, where
$\rho_0$ is the nuclear-saturation density \cite{Weih2019}), EOSs that
allow for a PT have received increasing attention in the recent past
\cite{Most2018,Montana2018,AlvarezCastillo2018,Christian2019,Most2019c}.

However, the lack of the detection of a post-merger signal from merging
BNSs leaves the issue of the occurrence of a PT still unsettled, but it
also motivates all those theoretical studies that can highlight the
various manifestations in which this process will reveal itself
\cite{Oechslin:2004,Most2018b,Bauswein2019}. In this Letter, we
introduce a novel signature in which a PT can be detected from the
postmerger gravitational-wave signal of a BNS system. This new
signature, besides extending and completing our understanding of the
occurrence of a PT in BNS mergers, also promises to be the signature
that, better than those considered so far, will signal the production of
quark matter in the present Universe.

\begin{figure} [t!] 
	\includegraphics[width=1.0\columnwidth]{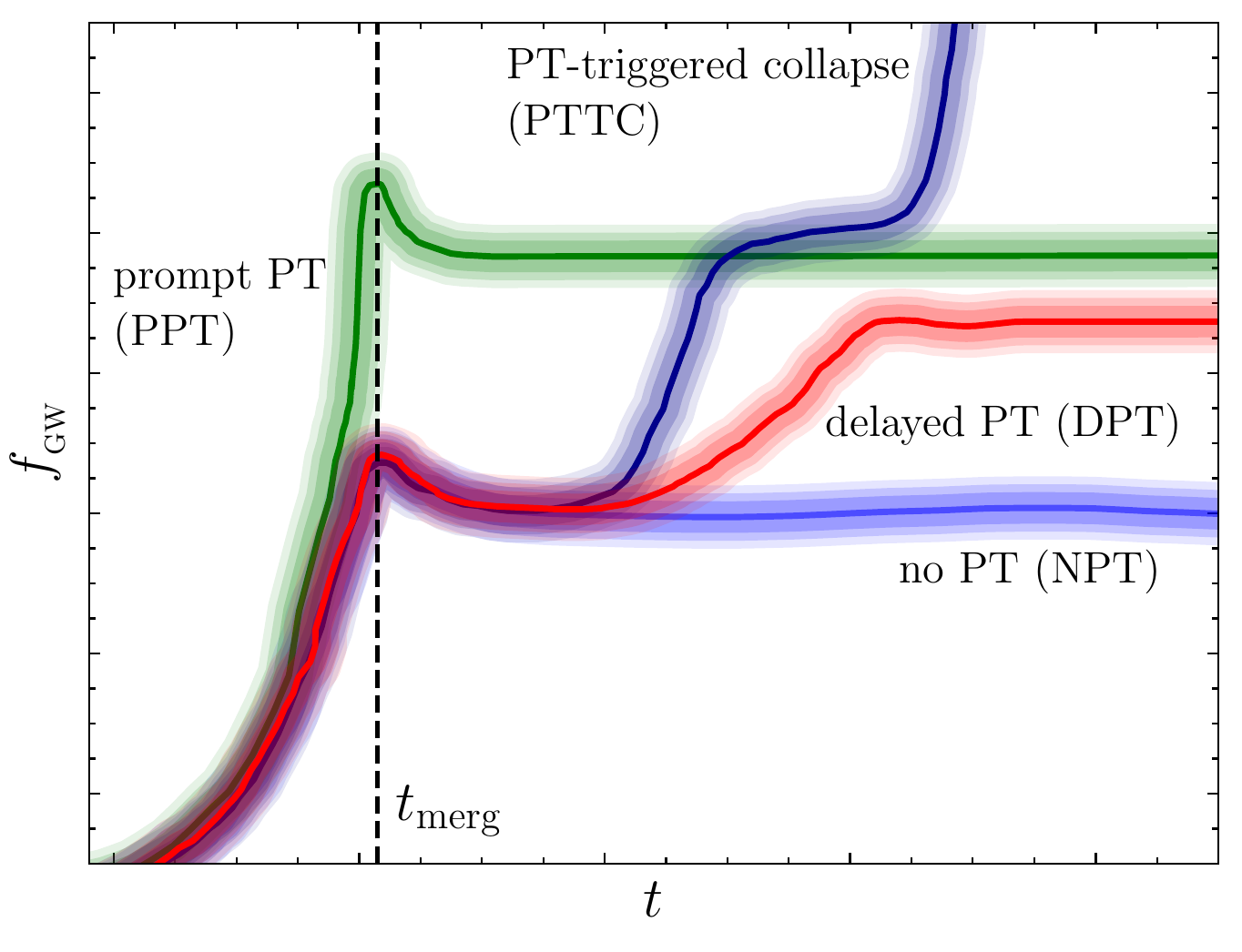}
        \caption{Schematic overview of the instantaneous characteristic
          gravitational-wave frequency and how its evolution can be used
          to classify the different scenarios associated with a
          hadron-quark PT.}
        \label{fig:schematic}
\end{figure}

Before discussing the results of our simulations, it is useful to
describe on general grounds the different manifestations in which a PT
from hadronic to quark matter can take place in a BNS merger. These
manifestations are best understood when looking at the instantaneous
\textit{characteristic frequency} of the gravitational-wave signal,
$f_{_{\rm{GW}}}$, as normally shown in spectrograms. To this scope,
Fig. \ref{fig:schematic} shows schematically the evolution of
$f_{_{\rm{GW}}}$ and identifies four different scenarios:\smallskip\\
\textbullet~\textit{No phase transition} (NPT; light-blue line): This is
the standard case considered so far in BNS simulations, where no PT sets
in after merger and the stars are always fully hadronic (see, \eg
\cite{Baiotti2016,Paschalidis2016} for some reviews).\\
\textbullet~\textit{Prompt phase transition} (PPT; green line): The PT
sets in right after the merger. As in the NPT-scenario, the dominant
frequency settles down to a constant value which is significantly higher
than it would be in the absence of a PT and can violate universal
relations
\cite{Bauswein2011,Takami:2014,Takami2015,Bernuzzi2015a,Rezzolla2016,Kawaguchi2018},
thus providing a signature of the PT \cite{Bauswein2019}.\\
\textbullet~\textit{Phase-transition triggered collapse} (PTTC; dark-blue
line): After the merger, the remnant's core does not immediately undergo
the PT, which sets in later on, when the density in the remnant's core
reaches a critical value. The prompt softening of the EOS and the
consequent increase in density leads to the collapse to a black hole (BH). The
premature ringdown signal represents a signature of the occurred PT
\cite{Most2018b}.\\
\textbullet~\textit{Delayed phase transition} (DPT; light-red line):
Similar to the PTTC scenario, the PT sets in only some time after the
merger. In contrast to the PTTC scenario, the softening of the EOS does
not lead to a collapse, but to a metastable hypermassive hybrid star
(HMHS) emitting gravitational waves at higher frequencies. The presence
of two distinct and clear characteristic frequencies represents a strong
signature of the occurred PT. A somewhat similar mechanism is
found in core-collapse supernovae and can lead to the delayed
collapse of a proto neutron star to a BH once neutrinos have diffused
out and the pressure support has been reduced \cite{Prakash97}; a more
detailed discussion and comparison with the DPT scenario is presented
in the Supplemental Material \citep{supplemental}.\smallskip\\
Given the poor knowledge of the EOS of neutron stars, it is very
difficult -- and not our intention here -- to ascertain how likely any
of these scenarios is. However, some insight can be gained from the
analysis carried out in Ref. \cite{Most2018b}, where it was found that
when choosing a uniform prior for a piecewise polytropic
parametrization of the EOS, the occurrence of a strong PT is not very
common. More precisely, it was found that only $\sim 5\%$ of the EOSs
constructed (which were more than $10^7$), would actually yield a
PT. Furthermore, even when considering a stellar model that yields a PT
in the right density range, it is not trivial to assess whether it will
lead to a DPT, a PTTC or a PPT scenario. The outcome, in fact, depends
on the exact value where the PT sets in and at which density the
pure-quark phase is reached. In what follows we will discuss in detail
the new DPT scenario.

\noindent\emph{Methods and setup.~} For our simulations we make use of a
piecewise polytropic representation of the hadronic EOS \texttt{FSU2H}
introduced in Refs. \cite{Tolos2017a,Tolos2017b}. More specifically, we
modify this EOS by introducing a PT from hadronic to quark matter via a
Gibbs-like construction (\texttt{FSU2H-PT}). The simulations are
performed using \texttt{McLachlan} \cite{Brown2007b} for the spacetime
evolution, \texttt{WhiskyTHC} \cite{Radice2013b,Radice2013c} for solving
the equations of general-relativistic hydrodynamics, and \texttt{LORENE}
\cite{Gourgoulhon01} for setting up the initial irrotational binary
configurations. Details on this setup and the EOS can be found in the
Supplemental Material \citep{supplemental}.

With the exception of the EOS, the one described above is a
standard setup for simulating a BNS merger and sufficient to draw the
proof-of-concept scenario proposed here. For a more complete and
detailed picture, a fully temperature-dependent EOS \cite{Koeppel2019},
magnetic fields \cite{Most2019b}, and neutrino radiative transport
should be accounted for. However, we expect our results, which focus
only on the gravitational-wave signal produced over a few tens of
milliseconds after the merger, to remain unaltered when a more accurate
description of the microphysics is made.

Since a fully temperature-dependent EOS leading to a DPT scenario
is currently not available, we account for the additional shock heating
during the merger and postmerger phases by including thermal effects
via a ``hybrid EOS'', that is by adding an ideal-fluid thermal
component to the cold EOS \cite{Rezzolla_book:2013}. The total
pressure $p$ and the specific internal energy $\epsilon$ are therefore
composed of the cold part ($p_\mathrm{c},\epsilon_\mathrm{c}$) and a
``thermal'' ideal-fluid component ($p_\mathrm{th},\epsilon_\mathrm{th}$)
where
$p=p_\mathrm{c}+p_\mathrm{th}=K\rho^{\Gamma}+\rho\epsilon_\mathrm{th}\left(\Gamma_\mathrm{th}-1\right)$,
$\epsilon=\epsilon_\mathrm{c}+\epsilon_\mathrm{th}$, where $\rho$ is the
rest-mass density, $K$ the polytropic constant, and
$\Gamma_\mathrm{th}=1.75$.  The effective temperature obtained within
this ideal-gas approach can be roughly approximated as
$T=(m_{_{\mathrm{n}}}p_\mathrm{th})/(k_{_{\mathrm{B}}} \rho)$, where
$m_{_{\mathrm{n}}}$ is the nucleonic mass and $k_{_{\mathrm{B}}}$ the
Boltzmann constant.

\begin{figure} [t!]
  \includegraphics[width=1.0\columnwidth]{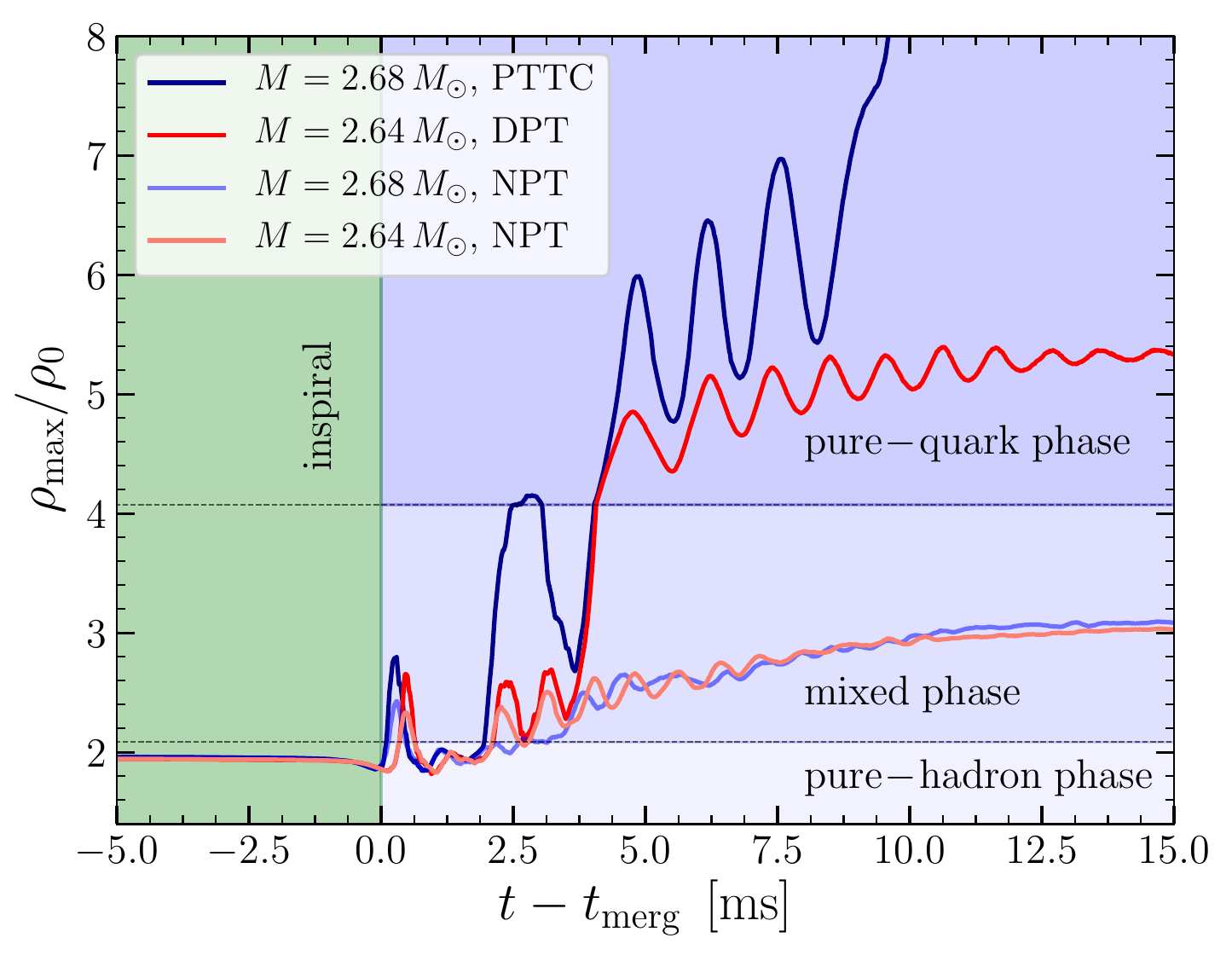}
  \caption{Evolution of the central rest-mass density for the four BNS
    configurations we have simulated. Blue-shaded regions mark the
    different phases of the EOS and apply to the DPT and PTTC scenarios
    only since the NPT binaries are always purely hadronic.}
  \label{fig:density}
\end{figure}
\begin{figure*}
  \includegraphics[width=1.0\textwidth]{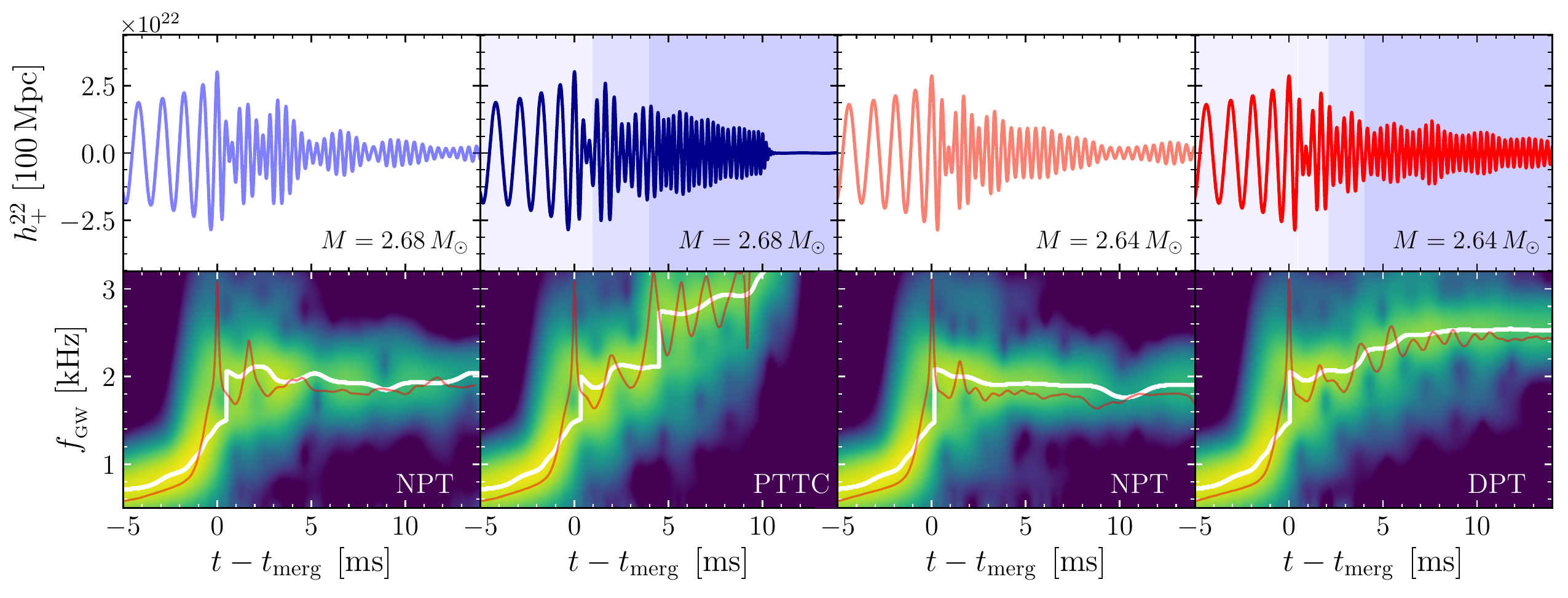}
  \caption{Strain $h_+^{22}$ (top) and its spectrogram (bottom) for the
    four BNSs considered. In the top panels the different shadings mark
    the times when the HMNS core enters the mixed and quark phases (\cf
    Fig. \ref{fig:density}); the NPT models are always purely
    hadronic. In the bottom panels, the white lines trace the maximum of
    the spectrograms, while the red lines show the instantaneous
    gravitational-wave frequency.}
  \label{fig:GW}
\end{figure*}

\noindent\emph{Results.~} Hereafter, we concentrate on two different and
representative equal-mass irrotational BNSs with $M=2.64M_\odot$ (low-mass) and
$2.68M_\odot$ (high-mass) for each of the two EOSs, \ie with
(\texttt{FSU2H-PT}) and without (\texttt{FSU2H}) a PT, for a total of four
simulations (the radii of the initial stars are $13.11$ and $13.13\,{\rm km}$,
respectively).  Figure \ref{fig:density} shows the evolution of the maximum
rest-mass density, $\rho_{\rm{max}}$, which is representative of the density
within the core of the merged object. As the evolution of BNSs without PT is
relatively well studied (see \cite{Baiotti2016,Paschalidis2016} for recent
reviews) and the matter in the NPT case is always purely hadronic we here focus
on describing the two simulations with PT and show the two simulations without
PT only for comparison as light-red and light-blue lines in Fig. \ref{fig:density}.

During the inspiral all models have densities below the onset of the PT,
so that the stars in this stage consist of purely hadronic matter. After
merger, the densities increase to values above the threshold for the
mixed phase of the cold EOS, but then quickly fall back below this
threshold. This local increase is simply due to the large compression
experienced by the stellar cores at the time of the merger and unless a
PT is triggered promptly (PPT case), $\rho_{\rm{max}}$ returns to the
typical values of the inspiral, so that the HMNS consists of purely
hadronic matter.

The large oscillations of the merged object cause the density in its core
to increase gradually and to reach values above the threshold of the
mixed phase. Under these conditions, the EOS softens significantly which,
in turn, amplifies the increase in density and leads to a considerable
conversion of hadrons to quarks. The latter are mostly concentrated in
the core of the merged object, which comprises $\sim20\%$ of the mass of
the binary, as already found in \cite{Most2018b,Bauswein2019}. The
subsequent evolution of the merged object will depend on the total mass
of the binary. More specifically, in the case of the high-mass binary
($M=2.68\,M_\odot$, dark-blue line in Fig. \ref{fig:density}), the
density experiences large oscillations, increasing considerably so as to
reach values $\rho_{\rm max}\approx (5-7)\,\rho_{0}$ and entering the
pure-quark phase. Once the PT is complete, the resulting EOS is
considerably softer in the mixed phase but also stiffer in the pure-quark
phase. However, the mass is sufficiently large so that gravity prevails
and hence the HMHS collapses rapidly to a BH. Despite the short time that
the quark core survives prior to collapse, the fact that the collapse is
still a direct cause of the PT essentially categorizes this evolution as
PTTC scenario which was already encountered in \cite{Most2018b}. Instead,
in the case of the low-mass binary ($M=2.64M_\odot$, red line in
Fig. \ref{fig:density}), the density also increases again entering the
pure-quark phase via large -- but comparatively smaller -- oscillations,
and thus marking the occurrence of the PT. However, quite differently
from the PTTC case, the new HMHS settles down to a new metastable
configuration with higher central density, $\rho_{\rm max}\approx
(4-5)\,\rho_{0}$, and thus becomes a steady emitter of gravitational
waves with a new and higher characteristic frequency. This new
equilibrium is again the result of the subtle balance between a softening
of the EOS in the mixed phase and a stiffening in the quark
core. However, in contrast with the high-mass binary that collapses to a
BH, for the low-mass binary, the stiffening is sufficient to
prevent the collapse and yield a metastable equilibrium.

\begin{figure*}
  \includegraphics[width=0.85\textwidth]{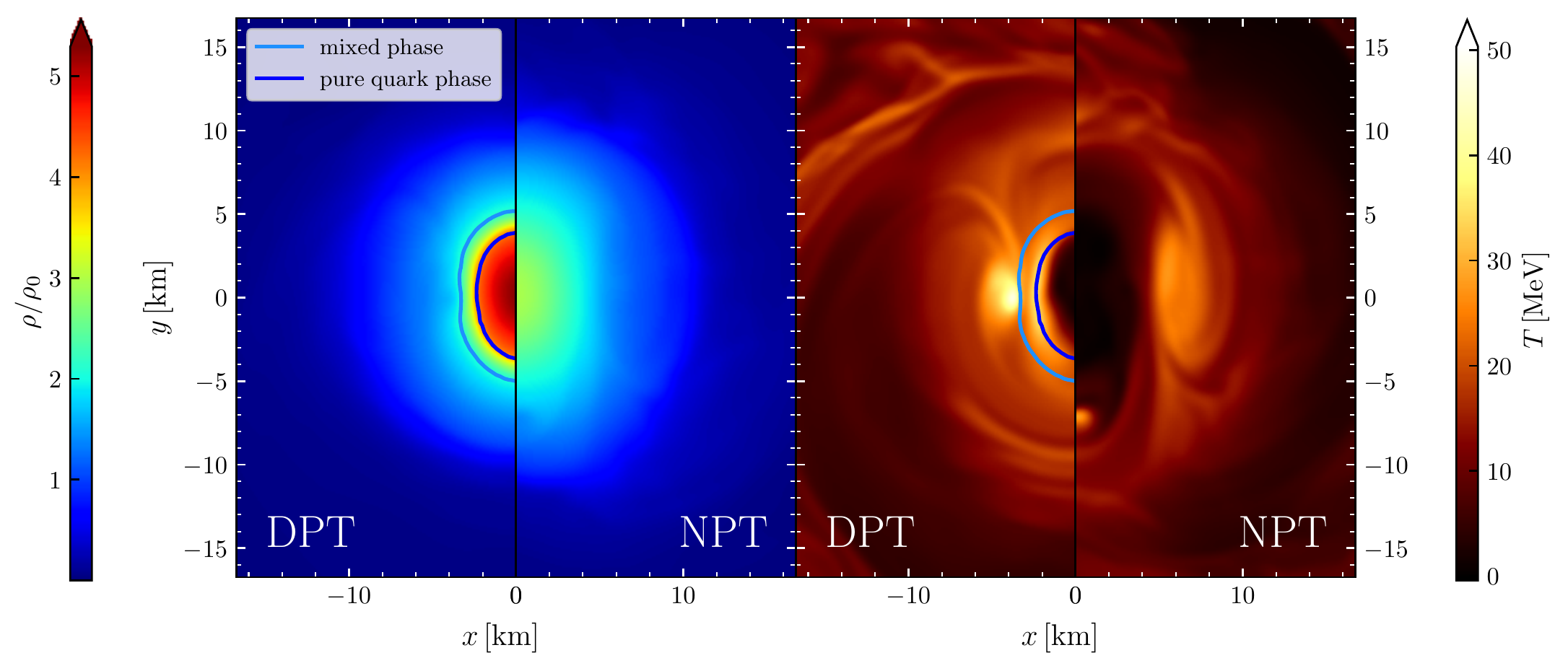}
  \caption{Comparison within a DPT scenario between the low-mass BNSs
    either undergoing a PT (left portions) or not (right portions). Shown
    are the rest-mass density (left panel) and temperature (right panel)
    at $t=12.85\, \rm{ms}$. The light-blue and dark-blue contours mark,
    respectively, the onset of the mixed phase and of the pure-quark
    phase, where the EOS is stiffened. Note that the ``hot-ring'' falls
    in the mixed-phase region, where the EOS is softened.}
  \label{fig:2Dsnaps}
\end{figure*}

Much of what was illustrated above for $\rho_{\rm{max}}$ is
faithfully reproduced by the gravitational-wave signal. This is shown in
Fig. \ref{fig:GW}, which reports in the top panels the $\ell=2=m$
component of the strain in the $+$ polarization, $h_+^{22}$, for the four
simulations (the blue shadings marking the various phases of matter use the 
same convention as in Fig. \ref{fig:density}). The
bottom panels, instead, show the corresponding spectrograms together with
the instantaneous maximum of the power spectral density (white line)
and the instantaneous gravitational-wave frequency (red line).

From left to right, the four panels show first the cases of the high-mass
binaries (without and with a PT) and subsequently those of the low-mass
binaries (without and with a PT). The properties of the waveforms in the
absence of a PT (NPT cases; first and third columns) have been described
many times in the literature
\cite{Bauswein2011,Takami:2014,Takami2015,Bernuzzi2015a,Rezzolla2016,Maione2017}
and basically exhibit the transition at merger from a chirping frequency
over to a triplet of peaks, \ie $f_1,f_2,f_3$ \cite{Takami:2014}, with
the highest and lowest frequencies disappearing after $\sim5\,{\rm ms}$
when the transient phase of the merger is complete (see \cite{Takami2015}
for a simple explanation). 

The second column reports instead the case of a PTTC and shows that after
the PT has started developing, a \emph{sudden} jump is measured in the
spectrogram, with the HMHS emitting gravitational waves at higher and
\emph{increasing} frequencies as the collapse proceeds and ends with the
formation of a BH. While a strong signature of the occurred PT,
this scenario may be difficult to detect because the emission at higher
frequencies is rapidly replaced by the incipient collapse to black
hole.

Finally, the fourth column reports the new case of a DPT (which is also
discussed in more detail in the Supplemental Material
\cite{supplemental}) and shows that after the PT has started developing,
a \emph{progressive} transition to a signal at higher frequencies can be
measured as the HMHS moves to a new metastable equilibrium characterised
by a pure-quark core, which is more compact and hence emitting
gravitational waves at higher frequencies. This transition can be easily
appreciated both in the time domain through the different shadings but,
more transparently, from the evolution of the maximum frequency in the
spectrogram. In the specific case considered here, this transition takes
place between $4$ and $7\,{\rm ms}$, but this is mostly the result of our
specific setup and could be made to take place at different times after
suitable tuning of the parameters that define the EOS. What is important
is that the postmerger signal in a DPT scenario will exhibit a
spectrogram moving from a quasistationary hadronic-star low-frequency
$f^{\textrm{h}}_{2}$, over to a new quasistationary hybrid-star
high-frequency $f^{\textrm{q}}_2$, where the relative difference $1 -
f^{\textrm{q}}_2/f^{\textrm{h}}_2\sim 25\%$ is found to be robust at
different resolutions. Because emission at constant frequencies leads to
distinctive peaks in the postmerger gravitational-wave spectrum
\cite{Takami:2014}, the signature of a DPT promises to be the strongest
of those considered so far (see also \cite{supplemental} for more
information). As is common in the postmerger signal, the optimal
signal-to-noise ratio will be obtained when the gravitational-wave cycles
emitted before and after the PT are comparable in number and for the
longest duration possible. Interestingly, the sudden occurrence
of a PT in the HMNS triggers a sizable gravitational-wave signal also
in the $\ell=2, m=1$ component of the strain; while this mode is
subdominant, with $|h^{21}_{+}|/|h^{22}_{+}| \lesssim 0.1$, the
detection of this additional mode after the PT will represent another
important signature of its occurrence \cite{supplemental}.

Finally, Fig. \ref{fig:2Dsnaps} shows a comparison of the rest-mass
density (left panel) and temperature (right panel) distributions on the
$(x,y)$ plane for the low-mass binary at a representative time during the
postmerger evolution ($t=12.85\,{\rm{ms}}$). In particular, for each
panel, the left portion refers to the binary experiencing the PT, while
the right portion illustrates the corresponding quantities in the absence
of a PT. It is interesting to note that the BNS undergoing the PT
exhibits a much denser core, with densities that are almost twice as
large as in the absence of a PT. Such a core occupies a considerable
fraction of the central regions of the HMHS, while the mixed phase is
concentrated on a rather thin shell of $\sim1\,{\rm{km}}$
thickness. Marked differences can be found also in the temperature, which
is considerably higher in the binary with a PT. In both cases, however,
the central region is comparatively colder and the largest values in
temperature are concentrated in two opposite hot spots, which later
evolve into a ringlike structure
\cite{Hanauske2016,Kastaun2017}. Furthermore, in the case of a PT, these
two hot spots on either side of the core appear together with a
``hot ring''.  While the two hot spots mark the highest temperatures for
the purely hadronic regions of the HMHS, the hot ring falls within the
mixed-phase region. Indeed, the formation of this hot-ring takes place a
few milliseconds after the core underwent the PT and was therefore not
visible in the simulations reported in \cite{Most2018b}.

\noindent\emph{Conclusions.~} Exploiting the recent advances in the
simulation of BNSs with an EOS that allows for a PT
\cite{Most2018b,Bauswein2019}, we have introduced the first
classification of the postmerger gravitational-wave signatures of the
occurrence of a PT. The picture that emerges from this classification was
completed by the discussion of a novel scenario, which we refer to as
that of a \emph{delayed phase transition} (DPT). In this scenario, the
softening of the EOS resulting from the PT does not lead to the rapid
collapse to a BH, but to a metastable HMHS emitting gravitational waves
at higher frequencies. As a result, the postmerger signal in a DPT will
be characterized by an initial quasistationary low-frequency
$f^{\textrm{h}}_2$ emission corresponding to a (mostly) hadronic HMNS,
which then increases -- over a timescale of a few milliseconds -- to
reach a new quasistationary hybrid-star high-frequency
$f^{\textrm{q}}_2$ emission corresponding to a HMHS with a significant
quark core. Since a postmerger gravitational-wave emission with marked
peaks is comparatively easier to characterize, the signature of the novel
DPT promises to be the strongest in the proposed classification 
and hence the optimal signature to witness the creation of quark matter in the
present Universe.

As a concluding remark, we note that our classification of the occurrence
of a PT refers to the postmerger only since this is when the critical
densities (and temperatures) for the onset of a PT are more likely to be
reached. However, depending on the mass and mass ratio in the binary
system, a PT might take place already during the inspiral (\eg in systems
comprising a hybrid and a hadronic star or two hybrid stars), thus
enriching the range of manifestations in which a PT may appear. We plan
to explore these scenarios in future studies.

\medskip\noindent It is a pleasure to thank V. Dexheimer, E. Most,
G. Monta\~na, J. Papenfort, H. St\"ocker, and L. Tolos for useful
discussions. LRW acknowledges support from HGS-HIRe. Support also comes
in part from ``PHAROS'', COST Action CA16214; LOEWE-Program in HIC for
FAIR; the ERC Synergy Grant ``BlackHoleCam: Imaging the Event Horizon of
Black Holes'' (Grant No. 610058). The simulations were performed on the
SuperMUC and SuperMUC-NG clusters at the LRZ in Garching, on the LOEWE
cluster in CSC in Frankfurt, and on the HazelHen cluster at the HLRS in
Stuttgart.


\bibliography{aeireferences,local}
\bibliographystyle{apsrev4-1}

\clearpage



\section*{Supplementary material}

## SUPPLEMENTAL MATERIAL

### Numerical setup and equation of state

For the hadronic phase of the cold EOS discussed in the main text, we have used the `FSU2H` relativistic mean-field model [1, 2]. Similar to the `Model-2` of Ref. [3], a first-order transition from hadronic to quark matter has been incorporated by assuming a moderate value of the surface tension of deconfined quark matter droplets within the mixed phase. At zero temperature, this mixed phase follows the Gibbs conditions [4] leading to hadron-quark pasta phase structures [5]. The softening of the EOS within the mixed phase and its stiffening in the pure quark-matter phase has been modeled by two additional pieces of the EOS, following a polytropic dependence. The overall cold EOS used within our numerical approach (`FSU2H-PT`) consists of a piecewise polytrope representation [6] of the hadronic FSU2H model, a soft mixed phase region which starts at $2.085\,\rho_0$ (polytropic index $\Gamma = 1.04$) and a stiff pure quark-matter part for densities above $4.072\,\rho_0$ (polytropic index $\Gamma = 5.1$). In order to ensure that the sound speed of deconfined quark matter is always below the speed of light, three additional piecewise polytropes have been added for ultra-high densities ($\rho/\rho_0 = 4.823, 4.969, 5.289; \Gamma = 4.7, 4.1, 3.1$). The specific value of the EOS results in a hybrid star model with a small twin-star branch belonging formally to the twin-star *Category III* (see Fig. 1 of the supplemental material). We should note that the EOS considered here is not necessarily the most realistic one and a different choice of EOS parameters would increase the range in mass between the DPT and PTTC scenarios. However, the EOS has been constructed mostly to highlight the impact that a PT would have on the gravitational-wave signal.

The simulations are performed by solving the coupled Einstein-hydrodynamic equations [6] implemented within the general framework of the `Einstein Toolkit` [7]. The spacetime is evolved by the thorn `McLachlan` [8], which implements the conformal and covariant CCZ4 formulation of the Einstein equations [9] using the same gauge conditions as in [10, 11]. The equations of general-relativistic hydrodynamics are solved by the `WhiskyTHC` code [12, 13] via a high-order finite-differencing scheme using the HLLE approximate Riemann solver, the MP5 reconstruction scheme and a positivity-preserving limiter. For the time integration, we use the method-of-lines with a third-order Runge-Kutta scheme and a CFL number of 0.15 [6].

The initial irrotational binary configuration is computed with the `LORENE` code [14], using an initial separation of $45\,\mathrm{km}$. The grid is handled by the thorn `Carpet` [15], which implements box-in-box adaptive mesh refinement. We use six refinement levels with the finest level having a resolution of $dx = 0.16\,M_\odot\,(\simeq 237\,\mathrm{m})$, while the outer boundary is placed at a distance of $\sim 1500\,\mathrm{km}$ (simulations at higher and smaller resolutions have also been performed yielding consistent results; for compactness, we will omit their discussion here).

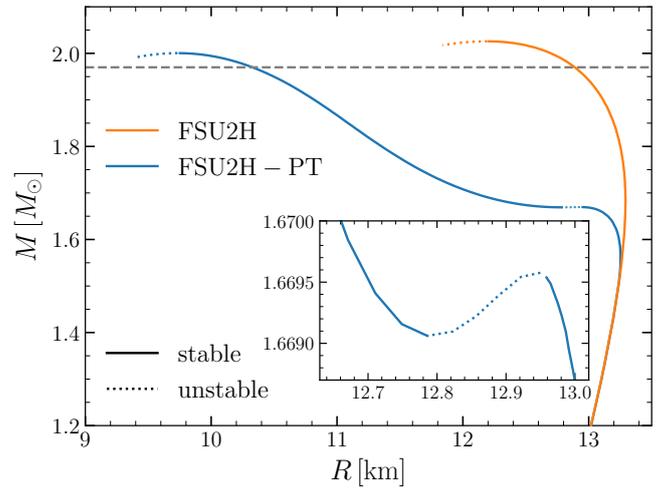

FIG. 1. Mass-radius relation for the purely hadronic EOS (`FSU2H`) and its modified version (`FSU2H-PT`). The latter shows a second stable (solid lines) branch after a small region of instability (dotted). The grey dashed line marks the limit of $1.97\,M_\odot$.

Furthermore, we reflect the grid across the $z = 0$ plane to reduce the computational cost.

### On the gravitational-wave signatures

Since the DPT scenario promises to deliver the clearest gravitational-wave signature among the cases considered, we here provide a few more details on its appearance. As already highlighted in the main paper, the signature is characterized by the transition of the postmerger frequency from a dominant frequency produced by the HMNS, i.e., $f_2^{\mathrm{h}}$, to a somewhat higher frequency $f_2^{\mathrm{q}}$ after the collapse to the metastable HMHS occurred. While this transition can be clearly seen in the right-most panel of Fig. 3 in the main text, Fig. 2 in this supplemental material shows that also the total spectrum of the gravitational-wave signal has a clear second peak corresponding to $f_2^{\mathrm{q}}$. More specifically, the left and right panels of Fig. 2 above report the total power spectral densities of models with $M = 2.64\,M_\odot$ corresponding either to no PT (left) or to a DPT (right). Clearly, the postmerger spectrum has only one major peak at $f_2^{\mathrm{h}}$ for the fully hadronic star, while it shows *also* a prominent $f_2^{\mathrm{q}}$ peak at higher frequencies in the case of a merger leading to a DPT. Such a "double-$f_2$" spectrum is obviously an important feature that will make a DPT easily recognisable.

As mentioned in the main text, the sudden occurrence of a PT in the HMNS, and the fact it will not proceed in a spherically symmetric fashion, will trigger a sizeable $\ell = 2, m = 1$ mode of the gravitational-wave signal. This is shown in Fig. 3 of the supplemental material for the low-mass runs. In particular, the top panel refers to the purely hadronic case (i.e., `FSU2H`), while the bottom one shows the $h_+^{21}$ strain in the case of an EOS with PT (i.e., `FSU2H-PT`). Both wave-

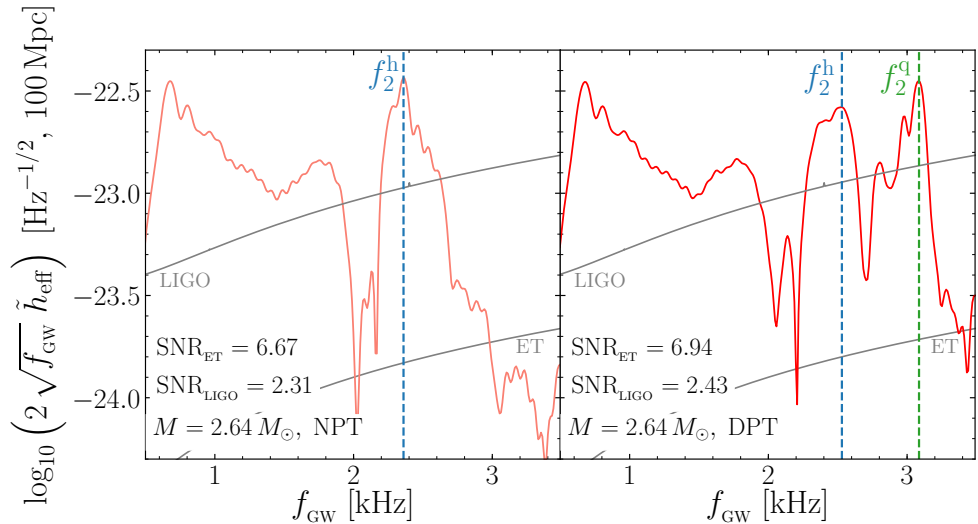

FIG. 2. Total spectrum of the ($\ell = 2 = m$)-mode of the gravitational-wave signal computed taking both the inspiral and the postmerger phase into account. The two panels compare the purely hadronic case (left) to the DPT scenario (right) and show that – in addition to the traditional $f_2^{\rm h}$ peak – there is a clear second peak at $f_2^{\rm q}$ in the latter case. Grey lines show the sensitivity curves of advanced LIGO and ET, respectively. The $f_2$-peaks are marked with dashed lines.

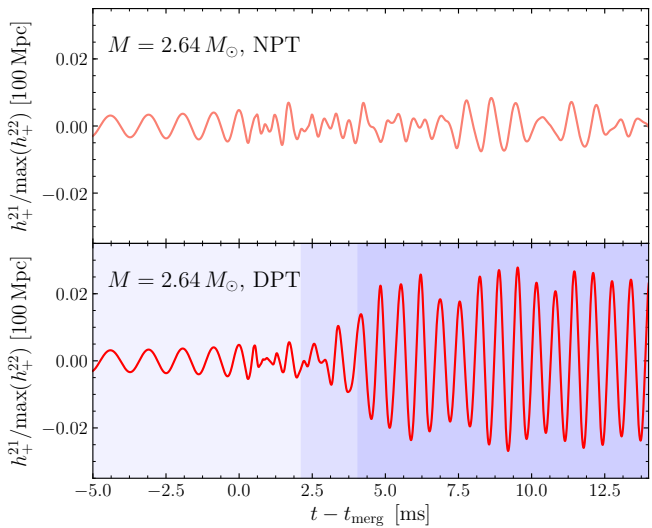

FIG. 3. $\ell = 2, m = 1$, "plus" polarization mode of the gravitational-wave signal for the case with $M = 2.64\,M_\odot$ with (bottom) and without (top) PT. The signals are normalized to the maximum of $h_+^{22}$ to highlight the relative importance. The shaded regions in the bottom panel mark the different phases of the EOS following the convention used in the main paper.

forms are normalised to the maximum value of the $h_+^{22}$ mode, which remains the dominant one and is about two orders of magnitude larger. Note that in the DPT case (bottom panel), $h_+^{21}$ has a sudden jump in amplitude as the PT starts to take place (cf. the colorcoding of the shaded areas) becoming more than three times the corresponding value before the merger. This increase is due to the fact that small asymmetries in the HMNS are drastically amplified when the remnant collapses to the HMHS state. This amplification then causes the dense and quark-rich core of the HMHS to oscillate causing an oscillation mode that combines with the $\ell = 2 = m$-mode caused by the rotation. Clearly, this does not happen for the fully hadronic case (top panel). Although this mode will become sizeable after the merger reaching amplitudes $|h_+^{21}|/|h_+^{22}| \lesssim 0.1$, the detection of this additional mode after the PT will represent another important signature of its occurrence.

### DPT-like scenarios in core-collapse supernovae

As discussed in the main text, the DPT scenario is characterized by a delayed onset of the PT during the HMNS phase. Other than in the PTTC case, the PT does not lead to a collapse to a BH, but rather to another metastable configuration, i.e., the HMHS. This scenario is similar to a mechanism found for supernovae and that is discussed in detail in [16]. We here compare these two mechanisms, showing the way in which they follow the same underlying physical processes, but also in which way they differ.

In the case of supernovae produced by stellar progenitors with sufficiently large mass, the collapse to a BH will happen directly after the core bounce or after a short phase of accretion, if the star consists of purely hadronic matter. On the other hand, if the EOS allows also for negatively charged non-leptonic components, i.e., hyperons, kaons or quarks, this immediate collapse can be delayed. This is due to the fact that such an EOS can support a larger maximum mass if the neutron star has neutrino-trapped matter as compared to a

neutrino-free neutron star. Such an environment of trapped neutrinos may be produced inside a proto-neutron star, which would then collapse to a BH only after the neutrinos have diffused out, thus reducing the pressure support. The proto-neutron star might then collapse to a BH after a diffusion timescale of the order of $\sim 10\,\mathrm{s}$ after core bounce [16].

While the maximum-mass neutron star that a EOS can support is a general criterion for determining whether a merger remnant promptly collapses to a BH or not – independently of whether a PT takes place – it is not the deciding factor in determining whether a PT-induced collapse will be halted by the production of a metastable HMHS or result in a BH. This distinction is rather connected to the presence of a twin-branch in the mass-radius plane, as shown in Fig. 1 of the supplemental material. Given the existence of such a branch, it is possible that the remnant will enter the unstable region, which will induce a collapse. This collapse, however, can then be halted when entering the second stable branch (where the EOS abruptly stiffens), thus producing the metastable HMHS. Ultimately, this HMHS might still collapse due to the loss of rotational energy via gravitational-wave emission, which will happen on timescales much longer than the dynamical timescales explored in simulations. In principle – and similar to the above described mechanism for supernovae – the collapse of the HMHS to a BH can also be triggered by the diffusion of neutrinos out of the HMHS. Exploring this process would, however, require much longer simulations that also include the treatment of radiative transport of neutrinos.

Finally, it should be noted that the process of the DPT and the consequent production of a HMHS is also similar to another mechanism found for core-collapse supernovae. In Ref. [17], in fact, it was shown that the collapse of the stellar core can be halted if the EOS includes a PT. This halting occurs at densities around $\sim 4-5\,\rho_0$ and is due to the sudden stiffening of the EOS in the pure-quark phase. In the DPT case explored here, a very similar mechanism is at work, so that at $\sim 3.5\,\mathrm{ms}$ after the merger, the HMNS starts to collapse but this collapse is halted as soon as the HMNS' core enters the pure quark phase (and thus turns into the HMHS), where the EOS suddenly stiffens.



\end{document}